\documentstyle[12pt]{article}
\newcommand{\be}{\begin{equation}}
\newcommand{\ee}{\end{equation}}
\newcommand{\al}{\mbox{$\alpha$}}
\newcommand{\s}{\mbox{$\sigma$}}

\newcommand{\bi}[1]{\bibitem{#1}}
\newcommand{\fr}[2]{\frac{#1}{#2}}

\newcommand{\gm}{\mbox{$\gamma_{\mu}$}}
\newcommand{\Pm}{\mbox{$P_{\mu}$}}
\newcommand{\Pn}{\mbox{$P_{\nu}$}}
\newcommand{\Pa}{\mbox{$P_{\alpha}$}}
\newcommand{\ph}{\mbox{$\hat{p}$}}
\newcommand{\Ph}{\mbox{$\hat{P}$}}
\newcommand{\Gh}{\mbox{$\hat{\Gamma}$}}
\newcommand{\qh}{\mbox{$\hat{q}$}}

\newcommand{\gn}{\mbox{$\gamma_{\nu}$}}

\newcommand{\GD}{\mbox{$\tilde{G}$}}
\newcommand{\gf}{\mbox{$\gamma_{5}$}}

\newcommand{\Tr}{\mbox{Tr}}

\newcommand{\vpl}[5]{ \put(#1,#2){\begin{picture}(80,20)
 \multiput(0,#3)(0,#4){#5}{\oval(#3,#3)[l]}
 \multiput(0,0)(0,#4){#5}{\oval(#3,#3)[r]}  \end{picture}}  }

\begin{document}

\normalsize
\begin{flushright}{BINP 94-22\\ February 1994}
\end{flushright}
\vspace{1.0cm}
\begin{center}{\Large \bf CP-odd effective gluonic Lagrangian in the
Kobayashi-Maskawa model}\\
\vspace{1.0cm}
{\bf  M.E. Pospelov\footnote{pospelov@inp.nsk.su}}\\

Budker Institute of Nuclear Physics, 630090 Novosibirsk
\vspace{4.0cm}
\end{center}

\begin{abstract}
Schwinger operator method is applied for studying CP-odd
pure gluonic effective Lagrangian in the Standard Model at three-loop level.
The induced $\theta$-term vanishes by the same reasons as EDMs of quark and
W-boson to two-loop approximation.  A simple way is found to demonstrate these
cancellations. All other terms of the effective Lagrangian acquire
non-vanishing contributions.  The effective operator of dimension six,
Weinberg operator, is calculated explicitly. The corresponding contribution
to the EDM of neutron is much smaller than that comes from large distances.

\end{abstract}
\newpage

\section{Introduction} The Kobayashi-Maskawa (KM) model looks now as the most
natural description of CP-violation. It describes properly CP-odd phenomena in
the decays of neutral $K$-mesons and predicts extremely tiny CP-odd effects in
the flavour-conserving processes. Though its predictions for the electric
dipole moments (EDM) of elementary particles are far beyond the present
experimental facilities, the corresponding theoretical investigations are of
certain methodological interest.

The subject of this work is the calculation of
the CP-odd effective gluonic Lagrangian which appears in the Standard Model as
a result
of integration over quark and W-boson modes. This Lagrangian can be naturally
expanded in the series of gluon field operators of increasing dimension:
\be
S_{eff}=\int d^4x{\cal L}_{eff}(x)=\int d^4x\left(c_1
g^2G^a_{\mu\nu}\GD^a_{\mu\nu}+
c_2g^3f^{abc}\GD^a_{\alpha\beta}G^b_{\beta\mu}G^c_{\mu\alpha}+...\right),
\label{eq:Seff}
\ee
where $g$ is the chromoelectric charge,
$\GD^a_{\alpha\beta}=1/2G^a_{\mu\nu}\epsilon_{\mu\nu\alpha\beta}$. It is
assumed that the characteristic loop momenta are much larger than the inverse
scale on which field fluctuations occur.
The first term in (\ref{eq:Seff}) represents the induced $\theta$-term,
perturbative contribution to the total $\theta$-term of the theory.  The
next operator of dimension 6 was introduced originally by Weinberg
\cite{Wein}. In different classes of models violating CP-symmetry this operator
may give an important contribution to the neutron electric dipole moment
\cite{BU1,BU2}.

The violation of CP-symmetry in Standard Model originates from the
complexity of the KM matrix. To lowest, quadratic order in the weak interaction
all CP-odd flavour-conserving amplitudes turn to zero trivially. The point is
that in this approximation those amplitudes depend only on the moduli squared
of elements of the KM matrix, so the result cannot contain the CP-violating
phase.

CP-odd objects may arise in the Standard Model in the  fourth order in
semi-week
constant. However, the cancellation of EDMs of a quark and W-boson in this
approximation is firmly established now \cite{Shab,KhP}. The only known
non-vanishing formfactor to this approximation is the magnetic quadrupole
moment of W-boson \cite{mqm}. The finite EDMs can be
obtained only after hard gluon radiative corrections are taken into account. We
shall prove that in the absence of QCD radiative corrections the same mechanism
leads to the cancellation of induced $\theta$-term. In contrary to the recent
claim that the Weinberg operator is zero to this approximation \cite{Booth}, we
find that all operators of dim$\geq$6 acquire non-vanishing values.

\section{Schwinger operator method for calculating CP-odd Lagrangian} We are
going over now to the direct calculation of a few first terms of
CP-odd effective gluonic Lagrangian
in the Standard Model to three-loop approximation. The general structure of
the diagrams which could contribute to the effect in that approximation is
shown on Fig.1, where the solid line represents a quark loop, waved lines -
W-bosons.

The CP-odd part of the loop flavour structure reads as:
\begin{eqnarray}
2i\tilde{\delta}[d(c(b-s)t-t(b-s)c+t(b-s)u-u(b-s)t+u(b-s)c-c(b-s)u)\nonumber\\
+s(c(d-b)t-t(b-s)c+t(d-b)u-u(d-b)t+u(d-b)c-c(d-b)u)\nonumber\\
+b(c(s-d)t-t(s-d)c+t(s-d)u-u(s-d)t+u(s-d)c-c(s-d)u)]
\label{eq:fs}
\end{eqnarray}
For the KM matrix we use the standard parameterization of Ref.\cite{ok}
where the CP-odd invariant is
\be
\tilde{\delta}=\sin \delta c_1c_2c_3s_1^2s_2s_3.
\ee
The letters $u,\;d,\;s,\;c,\;b,\;t$ denote here the Green's functions of
the corresponding quarks. Each product of four quark propagators allows for
cyclic permutations of the kind
\[
udcs=dcsu=csud=sudc.
\]

Further considerations are based on the operator Schwinger method
\cite{S1} successfully extended on the QCD case by Novicov, Shifman,
Vainshtein and Zhakharov \cite{NSVZ}. It allows one to minimize the set of
calculations in introducing the operator $\hat{P}$:
\be
\langle x|\hat{P}|y\rangle=\langle
x|i\hat{D}|y\rangle=\gm(i\fr{\partial}{\partial
x_{\mu}}+g\fr{\lambda^c}{2}A_{\mu}^c(x))\delta^4(x-y),
\label{eq:Phat}
\ee
where $A_{\mu}^c(x)$ is the external gluonic field. Then the quark propagator
taken in the background gluonic field reads as:
\be
\langle vac|Tq^a(x)\bar{q}^b(y)|vac\rangle=\langle
x,a|i(\hat{P}-m)^{-1}|y,b\rangle=\langle
x,a|(\hat{P}-m)\fr{i}{P^2+ig/2(G\s)-m^2}|y,b\rangle.
\label{eq:gf}
\ee
The external field strength appears as a result of commutation of two $P$:
\be
[\Pm,\Pn]=iG^a_{\mu\nu}\fr{\lambda^c}{2}\equiv igG_{\mu\nu}
\label{eq:comm}
\ee
We assume also that the field has no source and satisfies classical equations
of motion\footnote{As far as we are interested in pure gluonic operators we can
omit quark currents in the r.h.s. of (\ref{eq:EM})}:
\be
D_{\mu}G_{\mu\nu}=0;\; D_{\mu}\GD_{\mu\nu}=0.
\label{eq:EM}
\ee

The general outline of our calculation is following. Using the specific
property of the flavour summation (\ref{eq:fs}), namely the antisymmetrization
in the masses of opposite fermionic lines, we rewrite the general expression
for the CP-odd amplitude in KM model via some commutators of functions
depending on
$P$. It gives us some powers of external field and its derivatives in the
nominator. If the explicit dimension of this part of expression is equal to
the dimension of operator of interest, we can forget about further
non-commutativity of $P$s in other parts and put $A=0$.

The generic formula for the effective action up to some renormalization
terms which will be discussed later looks as
\be
S_{eff}=-i\sum_{flavour}\Tr\left[\fr{\Ph+m_1}{\Ph\Ph-m_1^2}\Gh(m_2^2)\fr{1+\gf}{2}
\fr{\Ph+m_3}{\Ph\Ph-m_3^2}\Gh(m_4^2)\fr{1+\gf}{2}\right],
\label{eq:master}
\ee
where the sum over quark's masses $m_1,\,m_3$ and $m_2,\,m_4$ should be
performed according (\ref{eq:fs}). The $\Tr$ operation in this equation means
the
trace in colour, Lorentz and coordinate spaces:
\be
\Tr[...]\equiv \Tr_{L+C}\int d^4x\langle x|\,...\,|x\rangle,
\label{eq:Tr}
\ee
$\Gh$ denotes the mass operator of a quark in the external field:
\be
\Gh(m_i^2)=\fr{g_w^2}{2}\int\fr{d^4q}{(2\pi)^4}\gm\fr{1+\gf}{2}\fr{\qh-\Ph}
{(q-P)^2+ig/2(G\s)-m_i^2}
\gn\fr{1+\gf}{2}\fr{g_{\mu\nu}-q_{\mu}q_{\nu}/M^2}{q^2-M^2},
\label{eq:Gamma}
\ee
where $M$ is the mass of the W-boson, $g_w$ - semiweak charge. This mass
operator $\Gh$ allows for the expansion in series of external
field operators of increasing dimension with some invariant functions
depending on $P^2$ as coefficients. It will be shown later that for the
calculation of the Weinberg operator it is sufficient to keep in this expansion
only three first terms of {\em explicit} dimension 0, 2 and 3.:
\be
\Gh(m_i^2)=(\fr{1}{2}\{\Ph,F_i(P^2)\}+\fr{ig}{4}\{H_i(P^2),\{\Ph,G\s\}\}
+ gJ_i(P^2)D_{\alpha}G_{\mu\nu}\Pa\Pm\gn+...)\fr{1+\gf}{2}
\label{eq:Gexp}
\ee
Here $\{ ... \}$ denotes anticommutator. Some comments should be added at this
point. As far as we do not fix the concrete view of invariant functions, two
last terms in (\ref{eq:Gexp}) could be presented in the other form. The term of
dimension 2, for instance, could be written as $\{\Ph,\{H_i(P^2),G\s\}\}$. The
difference between these two forms, however, is an operator of dimension 4 and
it can be absorbed to the next term of this expansion. From the same reasons we
do not care about the antisymmetrization in the last term in (\ref{eq:Gexp}).
The only thing which should be checked is the absence of operators
$\{H_i(P^2),\{\Ph,\GD\s\}\}$ and $D_{\alpha}\GD_{\mu\nu}\Pa\Pm\gn$. This could
be made from the expression (\ref{eq:Gamma}) or in the framework of the usual
perturbative expansion.

The V - A structure of $\Gh$ cancels $m_1$ and $m_3$ in nominators of
(\ref{eq:master}). The only problem arises with renormalization terms which
violate pure left-handedness. Now we shall argue that renormalization never
contributes to the CP-odd effective Lagrangian to this approximation.

In the absence of external field the first term in (\ref{eq:Gexp}) reduces to
the usual unrenormalized mass operator in the V-A theory:
\be
\fr{1}{2}\{\Ph,F(P^2)\}\fr{1+\gf}{2}\left|_{A=0}= \ph\fr{1+\gf}{2}
f(p^2)\right..
\ee
The renormalization with respect to quark 1 from the left and quark 3 from the
right introduces into the mass operator the dependence
of external masses \cite{Shab,KhP}:
\be
\ph\fr{1+\gf}{2}\tilde{f}(p^2)-f_{13}[\ph \fr{1-\gf}{2}-m_1\fr{1-\gf}{2}
-m_3\fr{1+\gf}{2}],
\label{eq:rmo}
\ee
where $f_{13}$ and $\tilde{f}$ are expressed via the function $f$ and
masses $m_1,\;m_3$ as follows:
\be
\tilde{f}(p^2)=f(p^2)-\fr{m_1^2f_1-m_3^2f_3}{m_1^2-m_3^2},\;\;\;
f_{13}=\fr{m_1m_3(f_1-f_3)}{m_1^2-m_3^2};\;\;
f_i=f(p^2=m_i^2), \; \; i=1,\,3.
\label{eq:f13}
\ee
It is clear how to adopt this scheme for the formalism of the external
field. Now the first term of (\ref{eq:Gexp}) should be written in the following
form:
\be
(\fr{1}{2}\{\Ph,F(P^2)\}-\Ph\fr{m_1^2f_1-m_3^2f_3}{m_1^2-m_3^2})\fr{1+\gf}{2}
-\fr{f_{13}}{2}[\Ph  (1-\gf)-m_1(1-\gf)-m_3(1+\gf)],
\label{eq:RMO}
\ee
where constants $f_1,\,f_3$ and $f_{13}$ are determined in (\ref{eq:f13}).
The zeroth-order term of the perturbative expansion of this expression in
$A$ reproduces (\ref{eq:rmo}); the first-order term corresponds to the
vertex part renormalized in complying with the Ward identity. Other terms
appear to be free of renormalization.

Let us start our consideration of renormalization effects from the counterterms
of the "wrong" handedness, proportional to $f_{13}$. From flavour structure
(\ref{eq:fs}) of the fermionic loop it follows, in particular, that any
amplitude should be antisymmetrized in the masses $m_1$ and $m_3$ of the
opposite quark's lines. The cyclic permutation allowed under the trace symbol
simultaneously
leads to the antisymmetry with respect to interchange of $m_2$ and $m_4$. The
last property automatically means the vanishing of the amplitude
proportional to $f_{13}(m_2)f_{13}(m_4)$.  Thus, one of mass operators at least
possesses V - A structure. Then the contribution of counterterms, proportional
to $f_{13}$ can be easily evaluated to the form:

\begin{eqnarray}
\Tr\left[\fr{\Ph+m_1}{\Ph\Ph-m_1^2}\fr{f_{13}}{2}(\Ph
(1-\gf)-m_1(1-\gf)-m_3(1+\gf))
\fr{\Ph+m_3}{\Ph\Ph-m_3^2}\Gh(m_4^2)\fr{1+\gf}{2}\right]=\nonumber\\
-f_{13}m_1m_3\Tr\left[\fr{\Ph}{(p^2+i/2G\s-m_1^2)(p^2+i/2G\s-m_3^2)}
\Gh(m_4^2)\fr{1+\gf}{2}\right].
\end{eqnarray}
This expression is explicitly symmetric under the permutation
$m_1\,\leftrightarrow\,m_3$ and drops out from the answer. Therefore, only V
- A structures left in both operators $\Gh$ and our master formula reduces to
\be
S_{eff}=-i\sum_{flavour}\Tr\left[\fr{\Ph}{\Ph\Ph-m_1^2}\Gh(m_2^2)
\fr{\Ph}{\Ph\Ph-m_3^2}\Gh(m_4^2)\fr{1+\gf}{2}\right].
\ee
The procedure of the antisymmetrization in $m_1\,\leftrightarrow\,m_3$
implied at this point significantly simplifies the set of further calculations.
Indeed,
the $m_1$,$m_3$-dependent part of the amplitude can be easily
transformed as follows:
\begin{eqnarray}
\fr{\Ph}{\Ph\Ph-m_1^2}\Gh\fr{\Ph}{\Ph\Ph-m_3^2}
-(m_1\,\leftrightarrow\,m_3)=\fr{\Ph}{\Ph\Ph-m_1^2}\Gh
\fr{\Ph\Ph-m_1^2}{\Ph\Ph-m_1^2}\fr{\Ph}{\Ph\Ph-m_3^2}
-(m_1\,\leftrightarrow\,m_3)=\nonumber\\
\fr{\Ph}{\Ph\Ph-m_1^2}[\Gh,\Ph\Ph]\fr{\Ph}{(\Ph\Ph-m_1^2)(\Ph\Ph-m_3^2)}
-(m_1\,\leftrightarrow\,m_3)=
(m_1^2-m_3^2)\Ph\hat{S}_{13}[\Gh,\Ph\Ph]\hat{S}_{13}\Ph,\nonumber\\
\label{eq:anti}
\end{eqnarray}
where we have introduced the operator
$\hat{S}_{13}=(\Ph\Ph-m_1^2)^{-1}(\Ph\Ph-m_3^2)^{-1}$. It is easy to see that
the antisymmetrization in $m_1\leftrightarrow m_3$ performed in (\ref{eq:anti})
leads to the antisymmetry of the amplitude with respect to interchange of
indices
2 and 4. Indeed, using the cyclic permutation we get:
\be
(m_1^2-m_3^2)\Tr\left(\Ph\hat{S}_{13}[\Gh_2,\Ph\Ph]\hat{S}_{13}\Ph\Gh_4\fr{1+\gf}{2}\right)=
-(m_1^2-m_3^2)\Tr\left(\Ph\hat{S}_{13}[\Gh_4,\Ph\Ph]\hat{S}_{13}\Ph\Gh_2\fr{1+\gf}{2}\right)
\ee
The calculation of the
commutator in the expression (\ref{eq:anti}) is quite straightforward.  For
instance, this commutator with the rest of the first term of the mass operator
can be calculated in the following manner:

\begin{eqnarray}
\left[(\fr{1}{2}\{\Ph,F(P^2)\}-\Ph\fr{m_1^2f_1-m_3^2f_3}{m_1^2-m_3^2}),
\Ph\Ph\right]\fr{1+\gf}{2}=
\fr{1}{2}\{\Ph,[F(P^2),P^2-\fr{ig}{2}G\s]\}\fr{1+\gf}{2}\nonumber\\=
\fr{ig}{2}(\{\Pa\{F'\{\Pm,D_{\mu}\GD_{\alpha\beta}\}\}\}
-\fr{ig}{4}\{F'\{\Pm,D_{\alpha}D_{\mu}G_{\alpha\beta}\}\})
\gamma_{\beta}\fr{1+\gf}{2}+\,{\cal O}(dim\geq 5).
\label{eq:Fcomm}
\end{eqnarray}
 This
formula is obtained using the relation from the general operator
calculus:
\be
2[F(A),B]=\{F'(A),[A,B]\}-\fr{1}{2!}[F''(A),[A,[A,B]]]
+\fr{1}{3!}\{F'''(A),[A,[A,[A,B]]]\}-+...\,\,.
\ee
The result of the commutation (\ref{eq:Fcomm}) brings some important
consequences:\vspace{0.5cm}\newline
First, it cancels the rest of renormalization counterterms. This completes the
prove of the statement that the renormalization never contributes to the
CP-odd flavor diagonal amplitudes to the fourth order in semi-weak constant.
\vspace{0.5cm}\newline
Second, the minimal explicit dimension of the operator $[\Gh,\Ph\Ph]$ is 3
and it puts the limit on the number of terms in $\Gh$ that we have to
take into account. For the calculation of the Weinberg operator we can restrict
our considerations on the operators of explicit dimension 3 or less and
it justifies, in particular, the choice of $\Gh$ in the form
(\ref{eq:Gexp}).\vspace{0.5cm}\newline
Third, we have found the shortest way to demonstrate the absence of
EDMs of the quark and W-boson to this approximation. There is no doubt that all
considerations presented above can be extended on the case of the external
electromagnetic field.

Now we are in the right position for the calculation of the Weinberg operator
in KM model. It is convenient to classify all contributions by the combination
of invariant functions from the expansion of $\Gh$s.  There are five of them:
J-H, J-F, H-H, H-F, F-F. For the induced $\theta$-term the only possible
contribution may arise from F-F.

The simplest cases are J-H and F-H because they already possess dim=6.
Therefore, the further non-commutativity may only influence on the effective
operators of higher dimension. Thus, it is reasonable to make the
substitutions:
\be
P\longrightarrow p;\; \hat{S}_{13}\longrightarrow
\fr{1}{(p^2-m_1^2)(p^2-m_3^2)}
\ee
Then the trace over spacial variables is easily computable:
\begin{eqnarray}
S_{eff}=4g^2\int\!d^4x
\Tr_C[(D_{\mu}\GD_{\alpha\beta})(D_{\nu}G_{\lambda\beta})]
\times \nonumber\\ \sum_{flavour}\!\int\!\fr{d^4p}{(2\pi)^4}
\fr{(m_1^2-m_3^2)p_{\alpha}p_{\mu}p_{\nu}p_{\lambda}p^2
[J_4(F'_2-H_2)-J_2(F'_4-H_4)]}{(p^2-m_1^2)^2(p^2-m_3^2)^2},
\end{eqnarray}
The trivial average over the direction of $p$
produces three different field operators of the dimension 6. By virtue of
equations of motion for the external field (\ref{eq:EM}) they can be
transformed to the standard form of
Weinberg operator.
Finally, we get J-H and J-F contributions to the Weinberg operator in the form:
\begin{equation}
c_2=i\sum_{flavour}\!\int\!\fr{d^4p}{(2\pi)^4}
\fr{(m_1^2-m_3^2)p^6
[J_4(F'_2-H_2)-J_2(F'_4-H_4)]}{(p^2-m_1^2)^2(p^2-m_3^2)^2}
\label{eq:JHF}
\ee

The analysis of the H-H contribution is quite transparent. The corresponding
amplitude contains field operators of dimension 5 and 6. It means that we can
neglect the non-commutativity of $H_2,\,H_4$ with operators $\Ph$ and
$\hat{S}_{13}$ because it would bring an additional dimension 2.
Then the amplitude
of interest could be transformed to the form:
\be
\Tr(\{H_2,[\hat{O},\Ph\Ph]\}\{H_4,\hat{O}\})=
\fr{1}{2}\Tr(\{H_2,[\hat{O},\Ph\Ph]\}\{H_4,\hat{O}\})
-\{H_4,[\hat{O},\Ph\Ph]\}\{H_2,\hat{O}\}),
\label{eq:HH}
\ee
where $\hat{O}=ig/4\{\Ph,G\s\}$. It is
the matter of simple exercise to check that the expression (\ref{eq:HH})
vanishes identically.

We left with the F-F and H-F groups of contributions which minimal dimension
is 3. The
straightforward calculation is quite tedious because we have to take into
account $G\s$-dependence of $\hat{S}_{13}$, the non-commutativity of different
$P$, etc. However, we have found a simple argument to show these groups do
not contribute to the Weinberg operator at all.
In the amplitude of interest (F-F case)
\be
\Tr(\Ph\hat{S}_{13}[\{\Ph,F_2(P^2)\},\Ph\Ph]\hat{S}_{13}\Ph\{\Ph,F_4(P^2)\}),
\label{eq:FF}
\ee
we perform the formal expansion in $ig/2G\s$ of
$F_i(P^2)=F_i(P^2+ig/2G\s-ig/2G\s)$ around the "point" $P^2+ig/2G\s=\Ph\Ph$.
It is clear that only zeroth and first order terms of that expansion could be
taken into account when calculating Weinberg operator. This expansion can be
performed using another formula of the general operator calculus. Up to terms
linear in $B$, function $F(A+B)$ could be expanded in the following manner:
\be
2F(A+B)=2F(A)+\{F'(A),B\}+\fr{1}{2!}[F''(A),[A,B]]+...
\ee

In our case this expansion takes the form:
\be
F(\Ph\Ph-ig/2G\s)=F(\Ph\Ph)-\fr{1}{2}\{F'(\Ph\Ph),ig/2G\s\}+\fr{1}{4}
[F''(\Ph\Ph),[\Ph\Ph,-ig/2G\s]]+...
\ee
Only second term of this expansion is relevant in our consideration. Indeed,
the third term and other denoted here as ... have the dimension higher than 3
and will not contribute to the Weinberg operator. First terms of the expansion
of $F_2$ and $F_4$ drop out because they give a vanishing commutator with
$\Ph\Ph$. So, we left with the second terms only but their contribution
literally coincides with H-H case. Thus, the vanishing of H-F and F-F groups of
contributions follows from the simple substitution
$H_i\longrightarrow H_i-F'_i$ in equation (\ref{eq:HH}). Another basis of
invariant functions
depending on $\Ph\Ph$ in the expansion of $\Gh$
\be
\Gh_i=(\{\Ph,\tilde{F}_i(\Ph^2)\}+\fr{ig}{4}\{\tilde{H}_i(\Ph^2),\{\Ph,G\s\}\}
+ g\tilde{J}_i(\Ph^2)D_{\alpha}G_{\mu\nu}\Pa\Pm\gn+...)\fr{1+\gf}{2}.
\ee
would simplify our analysis. It reduces the number of possible combinations for
the calculation of the Weinberg operator to $\tilde{H}-\tilde{H}$ and
$\tilde{H}-\tilde{J}$.

We have shown that the KM-model does not induce $\theta$-term to three-loop
approximation. To the same accuracy the Weinberg operator acquires nonvanishing
contributions of the form (\ref{eq:JHF}).

\section{Weinberg operator in the KM model}
After convincing ourselves in the absence of the exact cancellation of the
Weinberg operator in three-loop approximation, we are going to find its value.
It is natural to consider all quark masses but $m_t$ small as compared to the
W-boson one $M$. Together with the quark mass hierarchy it allows one to
simplify the calculations considerably, restricting to those contributions to
the operator of interest which are of lowest order in the light quark masses.
Besides, it is also natural to single out the contributions with logarithms of
large mass ratios, e.g., $\log(m_t/m_c),\,\log(m_b/m_s),\,\log(M/m_b)$ etc.

All diagrams can be split into two types, depending on which
quarks, $U\;(u,\;c,\;t)$ or $D\;(d,\;s,\;b)$, flow inside the mass
operators. It is convenient to sum first of all over the flavours of the quarks
masses of which were denoted up to now as $m_1$ and $m_3$. For the two
types mentioned we get respectively:
\begin{eqnarray}
\sum  \fr{(m_1^2-m_3^2)}{(p^2-m_1^2)^2(p^2-m_3^2)^2}\longrightarrow
 \fr{-m_b^4m_s^2}{p^4(p^2-m_b^2)^2(p^2-m_s^2)^2} \nonumber\\
 \fr{-m_t^4m_c^2}{p^4(p^2-m_t^2)^2(p^2-m_c^2)^2}.
\label{eq:sum}
\end{eqnarray}
In expression (\ref{eq:sum}) we put $m_u=m_d=0$.  We can determine now the
characteristic momenta $p$. When quarks are arranged according to
the first line of formula (\ref{eq:sum}), integral (\ref{eq:JHF}) is
infrared divergent if one neglects the masses $m_s$ and $m_b$ in the
denominator. It means that the typical loop momenta contributing to the
effect are $p\sim m_b$ and it cancels two powers of $m_b$ in the nominator of
(\ref{eq:sum}). In the opposite case when $D$-quarks are inside the
mass operators, the typical momenta range is large: $p\sim M$.

The main problem arising at this point is the calculation of invariant
functions $F$, $H$ and $J$. This could be done by means of usual perturbative
expansion of formulae (\ref{eq:Gamma}) and (\ref{eq:Gexp}). Using the fixed
point gauge
\begin{eqnarray}
(z-x)_{\mu}A_{\mu}(z)=0\\ \nonumber
A_{\mu}(z)=\fr{1}{2}(z-x)_{\nu}G_{\nu\mu}(x)+\fr{1}{3}(z-x)_{\alpha}(z-x)_{\nu}
D_{\alpha}G_{\nu\mu}(x)+...\;.
\end{eqnarray}
after straightforward calculations we get the following set of equations:
\be
F(p^2)\ph=\tilde{F}(p^2)\ph=\fr{g_w^2}{2}\int\fr{d^4q}{(2\pi)^4}
\fr{-2\qh-\fr{1}{M^2}(\ph\qh\ph+q^2\qh-2q^2\ph)}{(q^2-m^2)[(p-q)^2-M^2]}
\label{eq:F}
\ee
\be
H(p^2)\GD_{\mu\nu}p_{\mu}\gn=(\tilde{H}+\tilde{F})(p^2)\GD_{\mu\nu}p_{\mu}\gn=
\fr{g_w^2}{2}\int\fr{d^4q}{(2\pi)^4}\fr{(\gm q_{\nu}-\fr{2}{M^2}\qh
p_{\mu}q_{\nu})\GD_{\mu\nu}}
{(q^2-m^2)^2[(p-q)^2-M^2]}
\label{eq:H}
\ee
\be
\label{eq:J}
J(p^2)D_{\alpha}G_{\nu\mu}p_{\nu}p_{\alpha}\gm=
\fr{g_w^2}{3}\int\fr{d^4q}{(2\pi)^4}\fr{(\gm -\fr{2}{M^2}\qh
p_{\mu})q_{\nu}q_{\alpha}D_{\alpha}G_{\mu\nu}}
{(q^2-m^2)^3[(p-q)^2-M^2]}+\fr{F''}{3}D_{\alpha}G_{\nu\mu}p_{\nu}p_{\alpha}\gm
\ee
The cubic divergence of integral in the expression for $F$ is irrelevant for
us because the combination $F'(m_i^2)-F'(m_j^2)$ presenting in our equations
is obviously finite.

Before taking the integrals in (\ref{eq:F}) - (\ref{eq:J}) it is reasonable to
determine the light
quark mass dependence of $F'$, $F''$, $H$ and $J$ and sum over flavours left.
Clearly, the summation $\sum[J_4(F'_2-H_2)-J_2(F'_4-H_4)]$ brings an additional
factor $m_c^2$ for $U$-quarks, so the total degree of suppression for this type
of
diagrams is ${\cal O}(m_b^2m_c^2m_s^2)$. For the D-type of
quark's arrangement it is essential that functions
$F'$ and $F''$ could be expanded as follows:
\be
F'(p^2,m^2)=F'(p^2,m^2=0)+\left.m^2\fr{dF'}{d(m^2)}\right|_{m=0}+...\,,
\label{eq:ef}
\ee
whereas $H$ and $J$ contain pieces proportional to $m^2\log m^2$ (see
the Ref.\cite{mqm} for the details). In the sum over D-flavours these
logarithms prevent the cancellation of terms $\sim m_b^2m_s^2$
which means that the group of diagrams with
D-quarks inside mass operators contribute to the Weinberg operator at the same
order ${\cal O}(m_b^2m_c^2m_s^2)$. Moreover, these logarithmic factors enhance
the contribution came from $D$-type of quark arrangement in comparison with
that
from $U$-type. Up to the last integral over $p^2$ the
corresponding contribution to the Weinberg operator to double logarithmic
accuracy reads as:
\be
c_2=-\fr{i\tilde{\delta}}{3}\left(\fr{g_w^2}{32}\right)^2
\fr{m_b^2m_c^2m_s^2}{M^4}\log(m_b^2/m_s^2) \!\int\!\fr{d^4p}{(2\pi)^4}
\fr{(p^2+M^2)m_t^4}{(p^2-M^2)^5(p^2-m_t^2)^2}\log\fr{|p^2-M^2|}{m_b^2}.
\ee
Performing trivial integration we get the final formula for the Weinberg
operator in the Kobayashi-Maskawa model:
\be
\fr{1}{1536\pi^6}\tilde{\delta}G_F^2\fr{m_b^2m_c^2m_s^2}{M^4}\log(m_b^2/m_s^2)
\log(M^2/m_b^2)I(m_t^2/M^2),
\label{eq:answer}
\ee
where the function $I$ is
\be
I(x)=\fr{x^2}{(x-1)^4}\log x\left(3+\fr{12}{x-1}+\fr{10}{(x-1)^2}\right)-
\fr{x}{(x-1)^2}\left(3+\fr{13}{x-1}+\fr{5}{(x-1)^2}+\fr{10}{(x-1)^3}\right).
\label{eg:I}
\ee
The Fermi constant is introduced in (\ref{eq:answer}) according the standard
notation $G_F=\sqrt{2}g^2/(8M^2)$.

\section{Discussion}

We have shown that the Kobayashi-Maskawa model generates CP-odd effective
gluonic Lagrangian to three-loop approximation starting from the operator of
dimension 6. The expression for the Weinberg operator
(\ref{eq:answer}) parametrically coincides with the effective magnetic
quadrupole moment of $W$-boson \cite{mqm} appearing in this model in the same
fourth order in semi-weak constant. The attempt to prove the exact cancellation
of the Weinberg operator using the external field technic \cite{Booth}
seems to be incorrect. The author of this work believes that
antisymmetrizations of the amplitude (\ref{eq:master}) in $m_1,\,m_3$ and
$m_2,\, m_4$ should be imposed independently and {\em both} procedures
increase the effective dimension. In contrary, we have shown they are
connected and the amplitude antisymmetrized in one pair is automatically
antisymmetric in another one.

There is nothing surprising in the fact that the contribution to the electric
dipole moment of neutron came from (\ref{eq:answer}) is tremendously
small. For $m_t\sim 2 M$ we get
\be
c_2\simeq 10^{-27}(1\,Gev)^{-2}
\label{eq:value}
\ee
Using the result of the work \cite{Khats} we can estimate the corresponding
contribution to the NEDM as $10^{-41}\,e\,cm$. It is far beyond both current
experimental
limit and theoretically predicted NEDM came from large distances \cite{KhZ} as
well. The extreme smallness of (\ref{eq:value}) reflects not only usual
parametric suppression of the effect but amazingly small numerical coefficient
as well.

Finally, we would like to discuss a possible value of the Weinberg operator at
four-loop level. One additional hard gluon loop brings a factor like
$\al_s/(3\pi)\sim 10^{-2}$ but now the operator of interest appears in another
order in light quark masses. As a result we could expect the corresponding
coefficient $c_2$ being much larger than its three-loop value \cite{BU1}. To
obtain an estimation for the Weinberg operator we use the approach developed
in the work \cite{Kh}. Believing that all quark masses are much smaller than
the mass of $W$-boson one can use four-fermion contact limit restricting on the
contributions to the effective Lagrangian of order $M^{-4}$. The only possible
structure of diagrams is presented on the Fig.2. (Dashed line here is the
gluon propagator). The induced
$\theta$-term appears from this graph in the order ${\cal
O}(\al_sG_F^2m_s^2m_c^2)$ \cite{Kh}.
The heaviest masses enters only under logarithms here. The estimation
for the Weinberg operator quoted in \cite{BU1}, ${\cal O}(\al_sG_F^2m_c^2)$,
looks strange now because the ratio $c_2/c_1$ is of order $1/m_s^2$. If it is
true it makes questionable the validity of the expansion (\ref{eq:Seff}) at the
four-loop level. Performing the same analysis we came, however, to the another
estimation of the effect. The typical expression for the corresponding
amplitude before the last integral over gluon momentum $k^2$ looks as:
\be
\al_sG_F^2\int dk^2
\fr{m_c^2}{k^2}\log(m_t^2/k^2)\log(k^2/m_c^2)\fr{m_s^2}{k^2}\log(m_b^2/k^2),
\ee
where the infrared divergence should be cut off at the scale of $m_c^2$.
Correspondingly, the four-loop contribution to the Weinberg operator is of
order ${\cal O}(\al_sG_F^2m_s^2)$ which is two order of magnitude smaller than
the estimation cited above.

I would like to thank I.B. Khriplovich for helpful discussions.

\newpage

\setlength{\unitlength}{1mm}

\begin{figure}
\begin{picture}(200,80)(0,0)

\put (20,50){\oval(30,15)}
\vpl{27}{43.2}{1.5}{3}{5}
\vpl{13}{43.2}{1.5}{3}{5}

\put(70,50){\circle{15.2}}
\put(84,50){\circle{15.2}}
\put(98,50){\circle{15.2}}

\put(77,50){\circle*{1}}
\put(91,50){\circle*{1}}

\put(70,57){\line(0,1){2}}
\put(98,57){\line(0,1){2}}
\put(71.5,61){\oval(3,3)[tl]}
\put(96.5,61){\oval(3,3)[tr]}
\multiput(73.5,62.5)(4,0){6}{\line(1,0){2}}

\put (20,30){\makebox(0,0){Fig. 1}}
\put (84,30){\makebox(0,0){Fig. 2}}

\end{picture}
\end{figure}

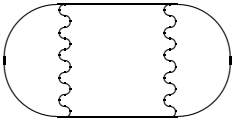
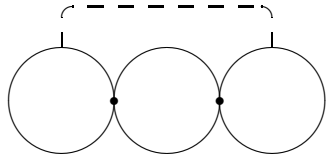
\end{document}